\documentclass[aps,prd,showpacs,twocolumn,groupedaddress]{revtex4}
\usepackage{dcolumn}
\usepackage{graphicx}
\usepackage{amsmath,amssymb}

\providecommand{\h}{h}

\begin{document}

\title{Mass spectrum of $1^{-+}$ exotic mesons from lattice QCD}

\author{M. S. Cook}
\email[]{mcook003@fiu.edu}
\author{H. R. Fiebig}
\email[]{fiebig@fiu.edu}
\thanks{This material is based on work supported by the U.S. National Science 
        Foundation under Grant No. PHY-0300065 and upon resources provided by the 
        Lattice Hadron Physics Collaboration through the SciDac program of the 
        U. S. Department of Energy.}
\affiliation{Department of Physics, Florida International University, \\ 
        Miami, Florida, USA 33199}

\date{\today}

\begin{abstract}
Time correlation functions of a hybrid exotic meson operator, with $J^{PC}=1^{-+}$,
generated in quenched lattice QCD are subjected to a (Bayesian) maximum entropy
analysis. Five distinct spectral levels are uncovered. Their extrapolation into
the physical pion mass region suggests a possible relationship to
experimentally known states $\pi_1(1400)$ and $\pi_1(1600)$, and also to a state in
the 2\,\,GeV region carrying the same quantum numbers.
\end{abstract}


\maketitle

\section{Introduction\label{intro}}

Quantum chromodynamics (QCD) in principle permits the existence of hadrons
containing both quarks and valence gluons. These are commonly known as hybrids.
For example, in the meson sector, the presence of explicit glue allows for states
with quantum numbers $J^{PC}=1^{-+}$. The standard quark model cannot describe
a meson with the above quantum numbers. Hadrons with this feature are called exotic.
Other exotic states are possible \cite{Barnes:2003vy}, however, in this article
we will solely consider the $1^{-+}$ hybrid exotic meson.

According to \cite{PDBook:2004} experimentally established $1^{-+}$ mesons
are $\pi(1400)$ and $\pi(1600)$.
Several experimental programs currently focus on studies of exotic hadrons
in general.

On the theoretical side
we expect the lattice formulation of QCD, being a first-principles approach,
to eventually yield the properties of hybrid exotics such as masses and decay widths.
Studies in this area have been but a few
\cite{Michael:2005tw,Hedditch:2005zf,Bernard:1997ib,McNeile:2002en,
Cook:2006tz,McNeile2006,McNeile:2002fh,McNeile:2002az}.
Results hint at masses in the 1.4~GeV region \cite{Hedditch:2005zf},
and also at a state close to 1.9~GeV \cite{Cook:2006tz,Bernard:1997ib}.

In the present article we examine the mass spectrum of $1^{-+}$ mesons based on
lattice QCD. We utilize the outcome of a numerical simulation, gauge field
configurations and quark propagators, which had been previously generated to
study hybrid exotic decay widths \cite{Cook:2006tz}.

The novel feature here is the use of Bayesian analysis, more precisely the maximum
entropy method (MEM) \cite{Jar96}. This analysis tool enables us to utilize the
entire range of lattice time slices, thus eliminating the necessity for a
subjective choice of a time interval defining a plateau in the effective mass function,
an approach almost universally used as an analysis method to extract hadron masses.
When all time slices are utilized, on a given lattice volume, the method delivers
unambiguous results, free from subjective choices of effective mass plateaus.
In addition, the MEM is quite naturally adapted to multiple states,
if present. We will particularly rely on this feature to uncover the
$1^{-+}$ hybrid exotic meson mass spectrum, up to somewhat above 2~GeV.

\section{\label{sec:lat}Lattice Simulation}

In the context of the decay width calculation \cite{Cook:2006tz} it was argued that
systematic errors would likely dominate the final result. For this reason the
lattice action and lattice sizes were selected to keep the numerical effort
moderate.
Although such a constraint would not seem to be necessary for our present aim of
`just' extracting spectral masses, we find that the lattices and quark propagators
generated in \cite{Cook:2006tz} give surprisingly relevant results.

Thus, we here present the $1^{-+}$ exotic hybrid mass spectrum based on the Wilson
gauge field action used with Wilson fermions, in the quenched approximation, on an
anisotropic lattice of size $12^3\times 24$. The bare aspect ratio is
$a_s/a_t=2$, with $a_s$ being the spatial and $a_t$ the temporal lattice constants.
Simulations were done at one value of a global gauge coupling $\beta=6.15$ and four
values of global hopping parameters $\kappa$, as they appear in Tab.~\ref{tab:results}
below. We refer the reader to \cite{Cook:2006tz} for a precise definition of these global
parameters.

In order to set the physical mass scale, standard local zero-momentum meson
operators of the form
\begin{equation}\label{mesonops} 
O_X(t)= 
\sum_{\vec{x}}\bar{d}_{a}(\vec{x}t) \Gamma u_{a}(\vec{x}t)\,,
\end{equation}
where $a$ is color, were employed for the pseudoscalar meson $X=\pi,\Gamma=\gamma_5$,
the vector meson $X=\rho,\Gamma=\gamma_i$,
and the $a_1$ meson $X=a_1,\Gamma=\gamma_i\gamma_5$. where 
$i=1,2,3$ are spatial directions.

For an operator coupling to $1^{-+}$ hybrid meson states, which we
collectively refer to with $\h$, we follow \cite{Bernard:1997ib} and use
\begin{equation}\label{hyopert} 
O_{\h}(t)= \sum_{1\le i < j \le 3}
\sum_{\vec{x}}\bar{d}_{a}(\vec{x}t) \gamma_{i} u_{b}(\vec{x}t) \,[F^{ab}_{ij}(\vec{x}t)- F^{\dagger ab}_{ij}(\vec{x}t)]\,.
\end{equation}
Here $a,b$ denote color indices and $F_{ij}(x)$ is a product of $SU(3)$ link matrices arranged in a
clover pattern
\begin{eqnarray}\label{gluonfields}         
\lefteqn{F_{\mu\nu}(x) = U_{\mu}(x)U_{\nu}(x+\hat{\mu})U^{\dagger}_{\mu}(x+\hat{\nu})U^{\dagger}_{\nu}(x)\notag}\\
&+& U_{\nu}(x)U^{\dagger}_{\mu}(x-\hat{\mu}+\hat{\nu})U^{\dagger}_{\nu}(x-\hat{\mu})U_{\mu}(x-\hat{\mu})\notag\\
&+& U^{\dagger}_{\mu}(x-\hat{\mu})U^{\dagger}_{\nu}(x-\hat{\mu}-\hat{\nu})U_{\mu}(x-\hat{\mu}-\hat{\nu})
U_{\nu}(x-\hat{\nu})\notag\\
&+& U^{\dagger}_{\nu}(x-\hat{\nu})U_{\mu}(x-\hat{\nu})U_{\nu}(x+\hat{\mu}-\hat{\nu})U^{\dagger}_{\mu}(x)\,,
\end{eqnarray}
which is used in the spatial planes only thus employing magnetic type gluons, in the rest frame.

The parity transformation ${\mathcal P}$ applied to (\ref{hyopert}) gives
${\mathcal P}O_{h}(t){\mathcal P}^{-1} = -O_{h}(t)$, as it should. This relation relies on
${\mathcal P} U_{i}(\vec{x},t) {\mathcal P}^{-1} = U_{-i}(-\vec{x},t)$
for $i=1,2,3$, using the notation of Ref.~\cite{Gupta:1998tz}.
At the quantum level the corresponding propagator respects exact (negative)
parity because the (quenched) gauge field action $S[U]$ is translationally invariant, and
satisfies ${\cal P} S[U] {\cal P}^{-1} = S[U]$.
However, in a numerical simulation those two properties are achieved only in the limit
of a large number of gauge field configurations, but are otherwise approximate.
Consequently, one should expect that the hybrid meson propagator is contaminated by states
of the wrong (positive) parity, albeit with a small amplitude.
We will revisit this point in the discussion of results.
 
With regard to charge conjugation ${\cal C}$ we encounter a similar situation.
For the purpose of discussion, changing $\h$ to a charge neutral operator $\h^{0}$
($\bar{d}u\rightarrow\bar{d}d,\bar{u}u$),
the proof of ${\mathcal C}O_{h^{0}}(t){\mathcal C}^{-1} = -O_{h^{0}}(t)$
relies on ${\mathcal C}F_{ij}(x){\mathcal C}^{-1}=F^\ast_{ij}(x)$, which comes from
${\mathcal C} U_{i}(x) {\mathcal C}^{-1} =U^{\ast}_{i}(x)$.
Again, at the quantum level the corresponding propagator respects
exact (positive) charge conjugation because of $S[U]=S[U^\ast]$.
In contrast to the parity case this relation is easily enforced in the simulation.
The configurations $[U]$ and $[U^\ast]$ are equally probable. Thus with each $[U]$ in the
ensemble of 200 configurations we also include $[U^\ast]$ and compute fermion propagators
for both of those. This strategy doubles the number of fermion propagators that need
to be computed to 400. However,
charge conjugation is now numerically exact, and this also appears to be the reason for an
observed noise reduction of simulation signals.

All meson operators $O_X(t)$, $X=\pi,\rho,a_1,\h$, are expanded by  
employing quark field smearing \cite{Alexandrou:1994ti} and gauge field fuzzing \cite{Alb87a}.

Smeared quark fields are constructed by spreading the original
field $\psi(x)$ over neighboring sites using a recursive procedure.
Define $\tilde{\psi}^{0}(x)=\psi(x)$, then
\begin{eqnarray}\label{psi1}
\tilde{\psi}^{k}(x)&=&\tilde{\psi}^{k-1}(x) + \sum^{3}_{m=1} \alpha_{m} U_{m}(x)
\tilde{\psi}^{k-1}(x+\hat{m})\notag\\
&&+\sum^{3}_{m=1} \alpha_{m}
U^{\dagger}_{m}(x-\hat{m}) \tilde{\psi}^{k-1}(x-\hat{m})\,,
\end{eqnarray}
and similarly for $\tilde{\bar{\psi}}^{k}(x)$, with $k=1\ldots K$.
Here $\alpha_{m}$ is a strength factor which controls the amount of the
smearing in spatial direction $m$.
To keep the magnitude of the resulting correlation functions numerically under control
it is useful to apply a rescaling factor after each smearing step. A suitable choice is
motivated by the following observation:
For the sake of this argument only, interpret $\psi,\bar{\psi}$ as Hilbert space operators
obeying standard anti-commutation relations, then we find
\begin{equation}
\{\tilde{\psi}^1(x),\tilde{\bar{\psi}}^{1}(x)\}=\openone (1+2\, \sum^{3}_{m=1} \alpha^{2}_{m})\,,
\end{equation}
where $\openone$ is the color-spin unit matrix. Thus, with
\begin{equation}
R=(1+2\,\sum^{3}_{m=1} \alpha^{2}_{m})^\frac{1}{2}
\end{equation}
the smeared quark fields are rescaled as
\begin{equation}\label{rescaleq}
\tilde{\psi}^{k}\leftarrow R^{-1}\tilde{\psi}^{k},\quad
\tilde{\bar{\psi}}^{k}\leftarrow R^{-1}\tilde{\bar{\psi}}^{k}\,.
\end{equation}
after each iteration step (\ref{psi1}).

Similarly gauge field fuzzing is implemented defining $\tilde{U}^{0}_{m}(x)=U_{m}(x)$, and
then adding a sum over staples
\begin{eqnarray}\label{iterU}
\lefteqn{\tilde{U}^{k}_{m}(x)=\tilde{U}^{k-1}_{m}(x)} & & \\
&+&\sum^{3}_{\underset{n\ne m}{n=1}}\rho_{n}\,[\tilde{U}^{k-1}_{n}(x)\tilde{U}^{k-1}_{m}(x+\hat{n})
\tilde{U}^{k-1\,\dagger}_{n}(x+\hat{m})\notag\\
&&+\tilde{U}^{k-1\,\dagger}_{n}(x-\hat{n})\tilde{U}^{k-1}_{m}(x-\hat{n})\tilde{U}^{k-1}_{n}(x+\hat{m}-
\hat{n})]\,,\notag
\end{eqnarray}
with fuzzing strengths $\rho_m$ in directions $m=1,2,3$.
Again, for numerical reasons, it is useful to rescale the fuzzy link variables. These are
no longer elements of SU(3). In particular,
\begin{equation}
\frac{1}{3}{\rm Tr}\biggl(\tilde{U}^{k\,\dagger}_{m}(x)\tilde{U}^{k}_{m}(x)\biggr)=S^2\,,
\end{equation}
which suggests the normalization prescription
\begin{equation}
\tilde{U}^{k}_{m}(x) \leftarrow S^{-1}\tilde{U}^{k}_{m}(x)
\end{equation}
enforcing a trace of one after each fuzzing step (\ref{iterU}).

Both normalization prescriptions are clearly invariant under gauge transformations.
We use common strength parameters $\alpha_m=\rho_m=2.5$ for all spatial directions.

The basis of the current simulation are meson operators $O_{X\{k\}}(t)$
built from (\ref{mesonops}) and (\ref{hyopert}) where $k=1\ldots K$ indicates that the
quark and gauge fields have been replaced with
smeared  and fuzzed fields (\ref{psi1}) and (\ref{iterU}), respectively, at a
common level $k$.
The corresponding, single meson, correlation functions
\begin{eqnarray}\label{corX}
\lefteqn{C_{X\{k,l\}}(t,t_0)=}&& \\
&&\langle O_{X\{k\}}(t) O_{X\{l\}}^{\dagger}(t_0)\rangle -\langle O_{X\{k\}}(t)\rangle\langle O_{X\{l\}}^{\dagger}(t_0)\rangle \notag
\end{eqnarray}
are the elements of a $K\times K$ matrix. We use $K=3$, and unsmeared fields ($k=0$) are ignored.
The same operators are used at both source and sink, so $C_{X}(t,t_0)$ is hermitian.
Due to our choice of flavor structure (\ref{hyopert}), the separable terms in (\ref{corX}) are zero.

\section{\label{sec:ana}Analysis}

The time evolution of the eigenvalues of $C_{X}(t,t_0)$ determines the mass spectrum for the meson $X$.
Toward extracting the latter a standard procedure would be to solve a generalized eigenvalue
problem \cite{Luscher:1990ck}
\begin{equation}\label{eigen}
C_X(t,t_0)\Psi(t)=C_X(t_1,t_0)\Psi(t)\Lambda(t)
\end{equation}
where $t_1$ is fixed, $\Psi(t)$ is a $K\times K$ matrix, its columns being the generalized 
eigenvectors, and $\Lambda(t)$ is real diagonal.
This approach is equivalent to a redefinition (linear combination) of the
operators that make up the correlator matrix. Its purpose is to improve the numerical
`quality' of effective mass function plateaus derived from the eigenvalues.

We shall not rely on effective mass functions here, but rather analyze correlator
eigenvalues by means of the maximum entropy method.
Within this framework, solving of the generalized eigenvalue problem (\ref{eigen}) offers no
advantage from a numerical point of view. 
Thus we will diagonalize $C_{X}(t,t_0)$ directly.

Many eigenvalues, behaving exponentially with $t$,
rapidly vanish into numerical noise. Conventional diagonalization methods do not work
well under those circumstances. Singular value decomposition (SVD), on the other hand,
is ideally suited to the task \cite{Golub96}. The SVD reads
\begin{equation}\label{svd}
C_{X}(t,t_0)=U_{X}(t,t_0)\,\Sigma_{X}(t,t_0)\,V^{\dagger}_{X}(t,t_0)\,,
\end{equation}
where $U_{X}(t,t_0)$ and $V_{X}(t,t_0)$ are unitary in our case~\footnote{This decomposition
easily generalizes to the case of a different number of operators at source and sink.}, and
$\Sigma_{X}(t,t_0)={\rm diag}(\sigma_{X;1}(t,t_0)\ldots\sigma_{X;K}(t,t_0))$ contains the
singular values satisfying $\sigma_{X;k}(t,t_0)\ge 0$.
The relation of the singular values $\sigma_{X;k}(t,t_0)$ to the eigenvalues
$\lambda_{X;k}(t,t_0)$ of $C_{X}(t,t_0)$ is given by the following theorem:
If $C_{X}(t,t_0)$ is non-degenerate and positive semi-definite then the set
of eigenvalues $\{\lambda_{X;k}(t,t_0)|k=1\ldots K\}$ and singular values
$\{\sigma_{X;k}(t,t_0)|k=1\ldots K\}$ are the same.
Proving this is a simple exercise in linear algebra, the columns of $U_{X}(t,t_0)$ and $V_{X}(t,t_0)$ are
eigenvectors which may differ by phase factors only. 
In case of a degeneracy, which is highly improbable in a numerical setting, the usual ambiguities
with respect to eigenvectors apply.
For simplicity we will refer to $\sigma_{X;k}(t,t_0)$ as eigenvalues.

\begin{figure}[t]
\includegraphics[width=80mm,angle=0]{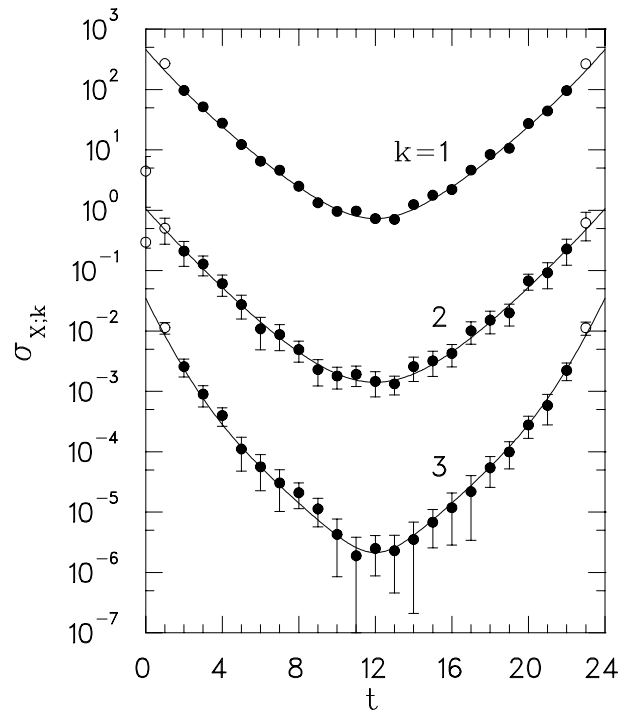}
\caption{\label{fig:corrfx}Eigenvalues of the $3\times 3$ correlation matrix
$C_{X}(t,t_0)$ for the hybrid meson $X=\h$ at the smallest pion mass.
Error bars are statistical, derived from a jackknife procedure \protect\cite{Efr79}.  
The lines are MEM fits as explained in the text. Open plot symbols indicate data
points not included in the analysis.}
\end{figure}

As an example, in Fig.~\ref{fig:corrfx} we show all eigenvalues of $C_{X}(t,t_0)$
for the hybrid meson operator, $X=\h$, computed at the smallest pion mass,
i.e. the largest $\kappa$ value. Diagonalization is done independently on each time slice.
Note that the eigenvalues are separated by almost three orders of magnitude.

Bayesian analysis methods have been considered for lattice QCD
\cite{Yamazaki:2001er,Asakawa:2000tr,Asakawa:2000pv,Lepage:2001ym,Fiebig:2001mr,Fiebig:2002sp},
but are not widely used despite obvious advantages.  
For instance, one may maximize the time slice fitting range. The time slices used for all
correlation functions in this work appear in Fig.~\ref{fig:corrfx} as filled plot symbols.

The Bayesian fitting problem can be set up and solved in a variety of ways. We here adopt
the framework described in detail in Ref.~\cite{Fiebig:2002sp}.
A brief summary will be useful:
The lattice provides correlation function data, say $\sigma(t)$, denoting any
set $\sigma_{X;k}(t,t_0)$ on the given time slice range $t=2\ldots 22$. Our data should be
well described by the model 
\begin{equation}\label{Fmodel}
F(\rho|t)=\int_{0}^{\infty}d\omega\,
\rho(\omega)\cosh(\omega(t-t_c))\,,
\end{equation}
where $t_c=12$ and $\rho(\omega)$ is a spectral density function.
From the Bayesian perspective the numbers $\rho(\omega)$ are interpreted as stochastic
variables. For these, employing Bayes' theorem, a conditional probability distribution
function $P[\rho\leftarrow \sigma]$ is constructed. Aside from inessential
normalization it has the form 
\begin{equation}\label{BayesT3}
{P}[\rho\leftarrow\sigma]\propto{P}[\sigma\leftarrow\rho]{P}[\rho]\,.
\end{equation}
Here ${P}[\sigma\leftarrow\rho]$ is known as the likelihood function. We construct it
from the $\chi^2$-distance between the data and the model
\begin{equation}\label{Chi2}
\chi^2=\sum_{t_1,t_2}
\left(\sigma(t_1)-F(\rho|t_1)\right)\Gamma^{-1}(t_1,t_2)
\left(\sigma(t_2)-F(\rho|t_2)\right)\,,
\end{equation}
with $\Gamma(t_1,t_2)$ being elements of the covariance matrix.
Then ${P}[\sigma\leftarrow\rho]=\exp(-\chi^2/2)$ is the choice for the likelihood function.
Bayesian inference uses information one might have about data, for example, a physical upper
limit on the spectral masses. Information like this is contained in ${P}[\rho]$, known
as the Bayesian prior. If no prior information is available, except say a physical mass
range, a suitable choice is based on the Shannon-Jaynes entropy \cite{Jar96}
\begin{equation}\label{Smem}
{S}=\int_{0}^{\Omega} d\omega\left(\rho(\omega)-m(\omega)
-\rho(\omega)\ln\frac{\rho(\omega)}{m(\omega)}\right)\,.
\end{equation}
Here $\Omega$ is the mentioned cutoff, and the function $m(\omega)$ is called the default model.
It provides a reference point for the spectral density in the sense that $S\leq 0$,
with $S=0$ if $\rho(\omega)=m(\omega)$.
Then ${P}[\rho]=\exp(\alpha S)$ is the choice for the Bayesian prior, introducing
a new parameter $\alpha$.
Hence the conditional probability (\ref{BayesT3}) for the spectral density $\rho$ becomes
\begin{equation}\label{Ppost}
{P}[\rho\leftarrow\sigma]\propto e^{-(\chi^2/2-\alpha S)}\,.
\end{equation}
The idea then is to find a spectral density function $\rho$ which maximizes
${P}[\rho\leftarrow\sigma]$, the posterior probability, at a fixed data set $\sigma$.
Bayesian inference within this framework is known as the maximum entropy method (MEM).
In \cite{Fiebig:2002sp} it was demonstrated that this problem can be solved by
simulated annealing or cooling. Consider the partition function
\begin{equation}\label{Zmem}
Z_W=\int [d\rho\/] e^{-\beta_W W[\rho\/]}\quad\mbox{with}\quad W[\rho]=\chi^2/2-\alpha S\,.
\end{equation}
It involves the generation of equilibrium configurations $[\rho]$ while
gradually increasing $\beta_W$ from an initially small value, following
some annealing schedule. For details we refer the reader to Ref.~\cite{Fiebig:2002sp}.
We here only mention that the resulting spectral density $\rho(\omega)$ is extremely
insensitive to both the default model $m(\omega)$ as well as to the entropy strength
parameter $\alpha$.

The numerical implementation of the above scheme requires discretization of the
$\omega$ integrals in (\ref{Fmodel}) and (\ref{Smem}). Choosing
$a_t\Omega=2.4$ for the spectral mass cutoff in (\ref{Smem}) we use $a_t\Delta\omega=0.05$.
The default model is a constant function $m(\omega)=10^{-6}a_t$ for all data sets $\sigma_{X;k}$,
and the entropy strength $\alpha$ is slightly adjusted, in each case, so as to
render the ratio of $\alpha S$ to $\chi^2/2$ between 0.1 and 0.01,
for the final $\rho(\omega)$.
Again, those parameter choices may be varied by several orders of magnitude without
significantly altering the resulting mass spectrum, see \cite{Fiebig:2002sp} for a discussion.

\begin{figure}[t]
\includegraphics[height=70mm,angle=90]{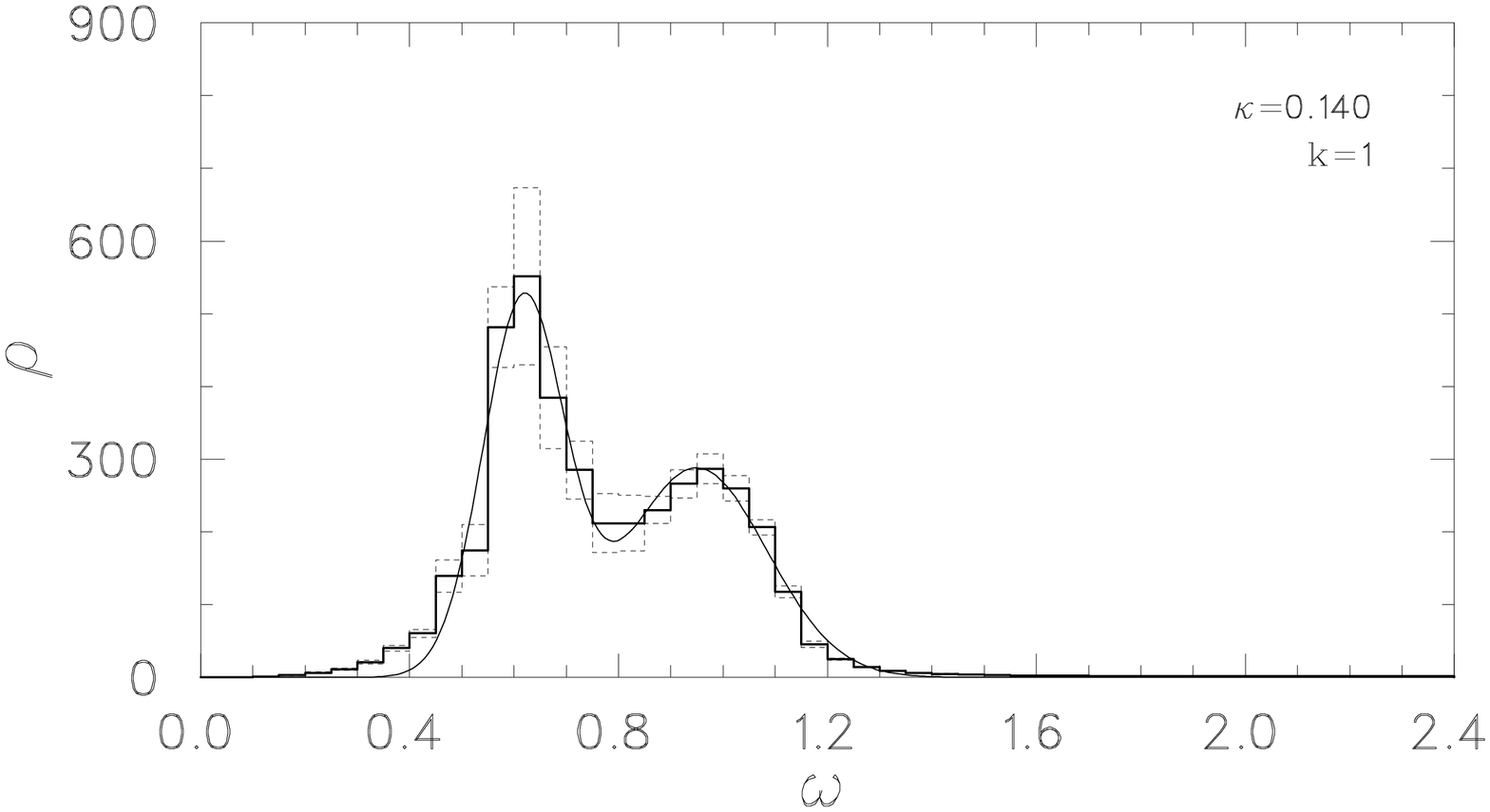}
\includegraphics[height=70mm,angle=90]{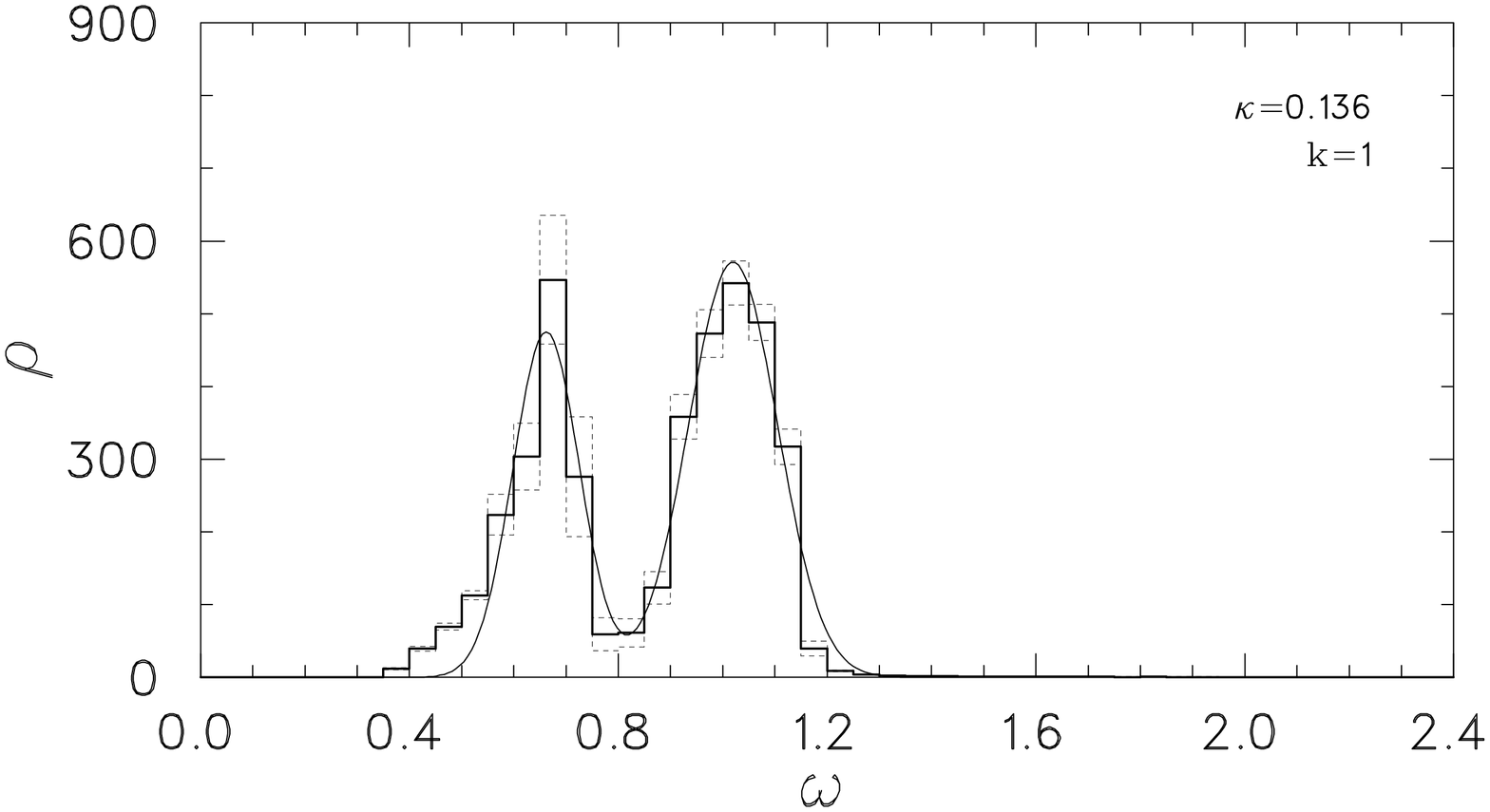}
\includegraphics[height=70mm,angle=90]{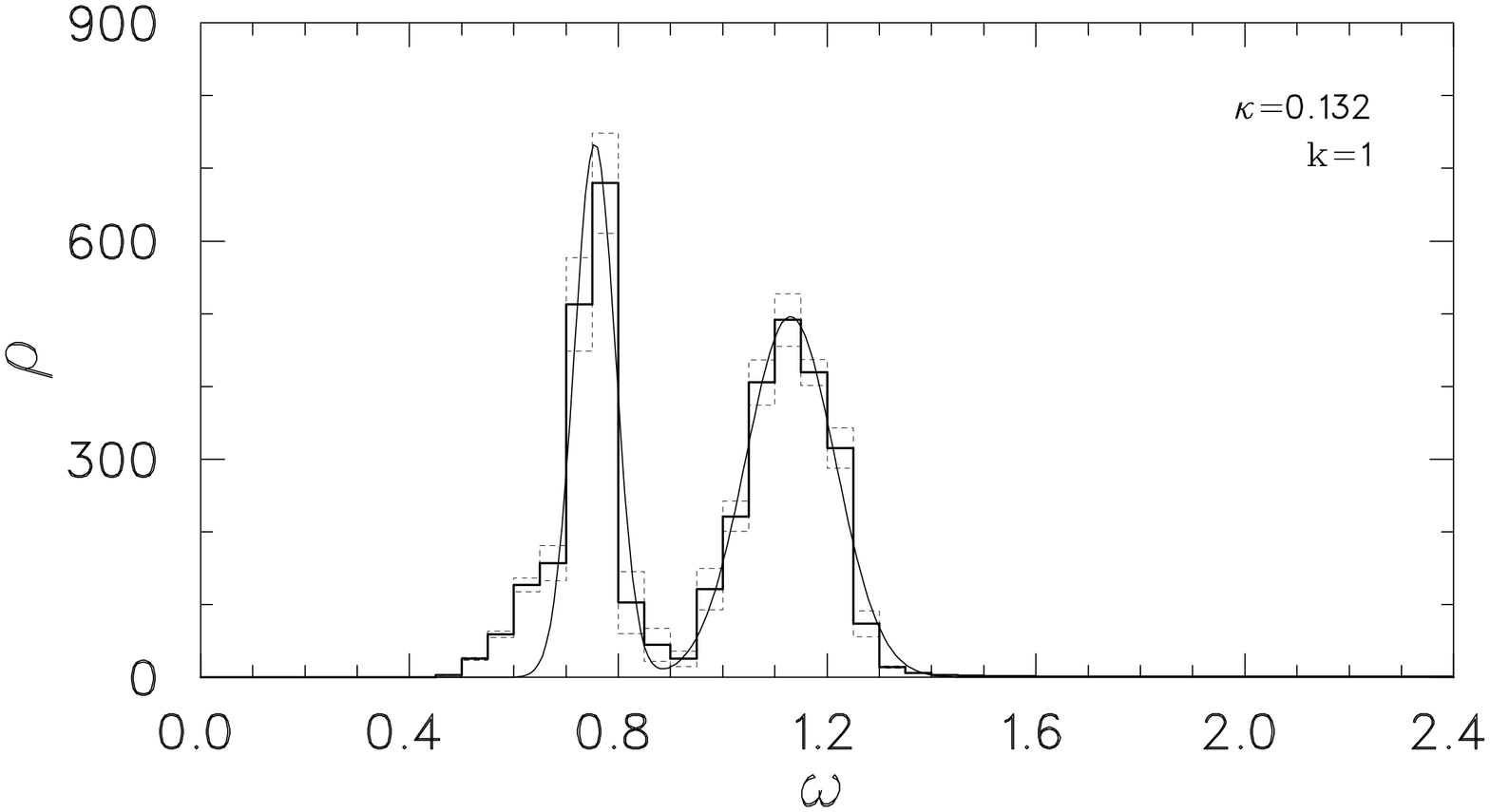}
\includegraphics[height=70mm,angle=90]{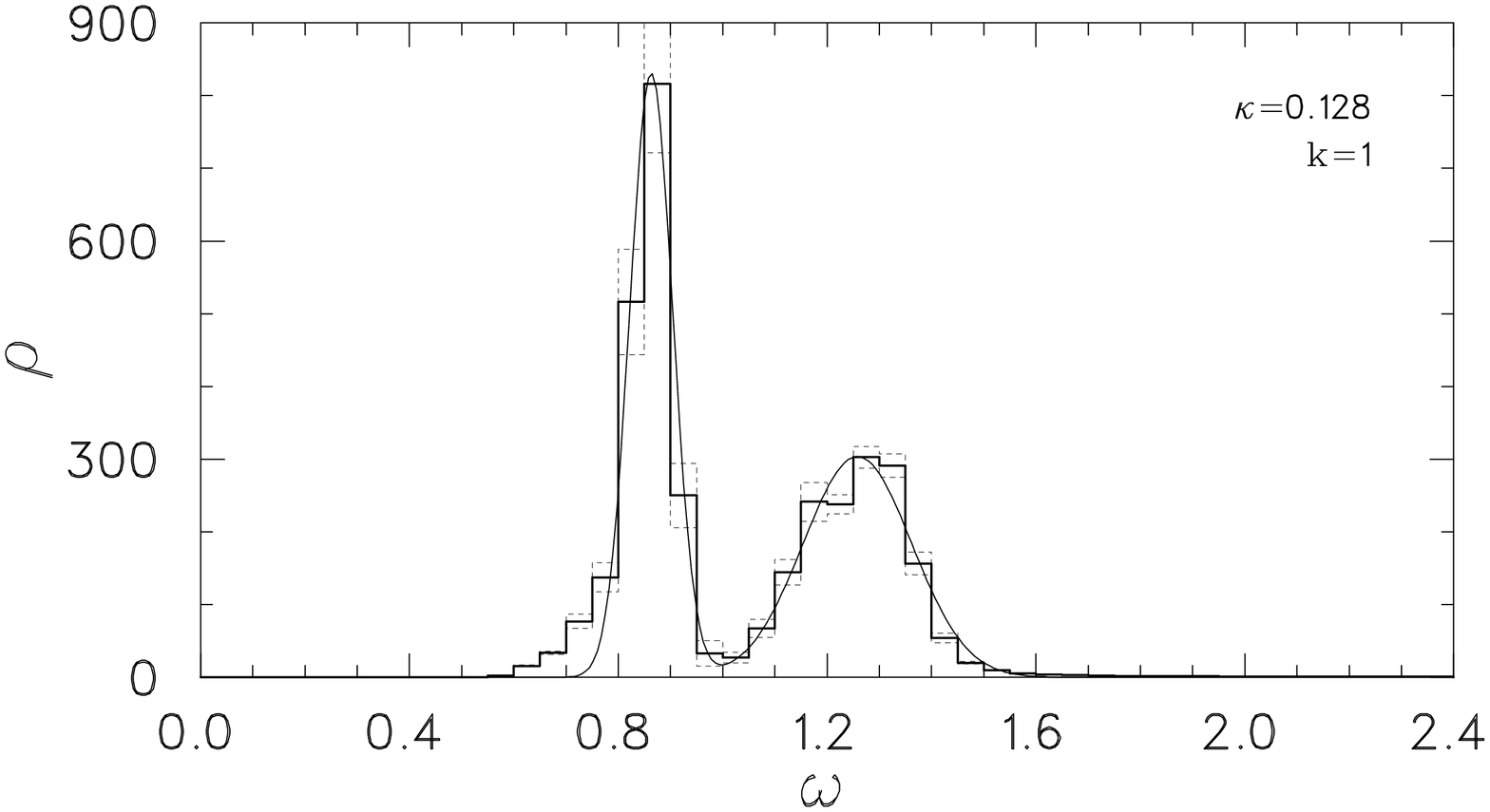}
\caption{\label{fig:memspecs1}Spectral density functions (thick solid histogram lines)
for the first (largest) eigenvalue set $\sigma_{\h,k}$, $k=1$,
of the hybrid meson correlation matrix.
The four panels correspond to increasing pion masses (decreasing hopping
parameters $\kappa$) from top to bottom.
The envelop histograms (thin dashed lines) indicate errors as explained in the text.
Smooth solid lines correspond to Gaussian fits.}
\end{figure}

\begin{figure}[t]
\includegraphics[height=70mm,angle=90]{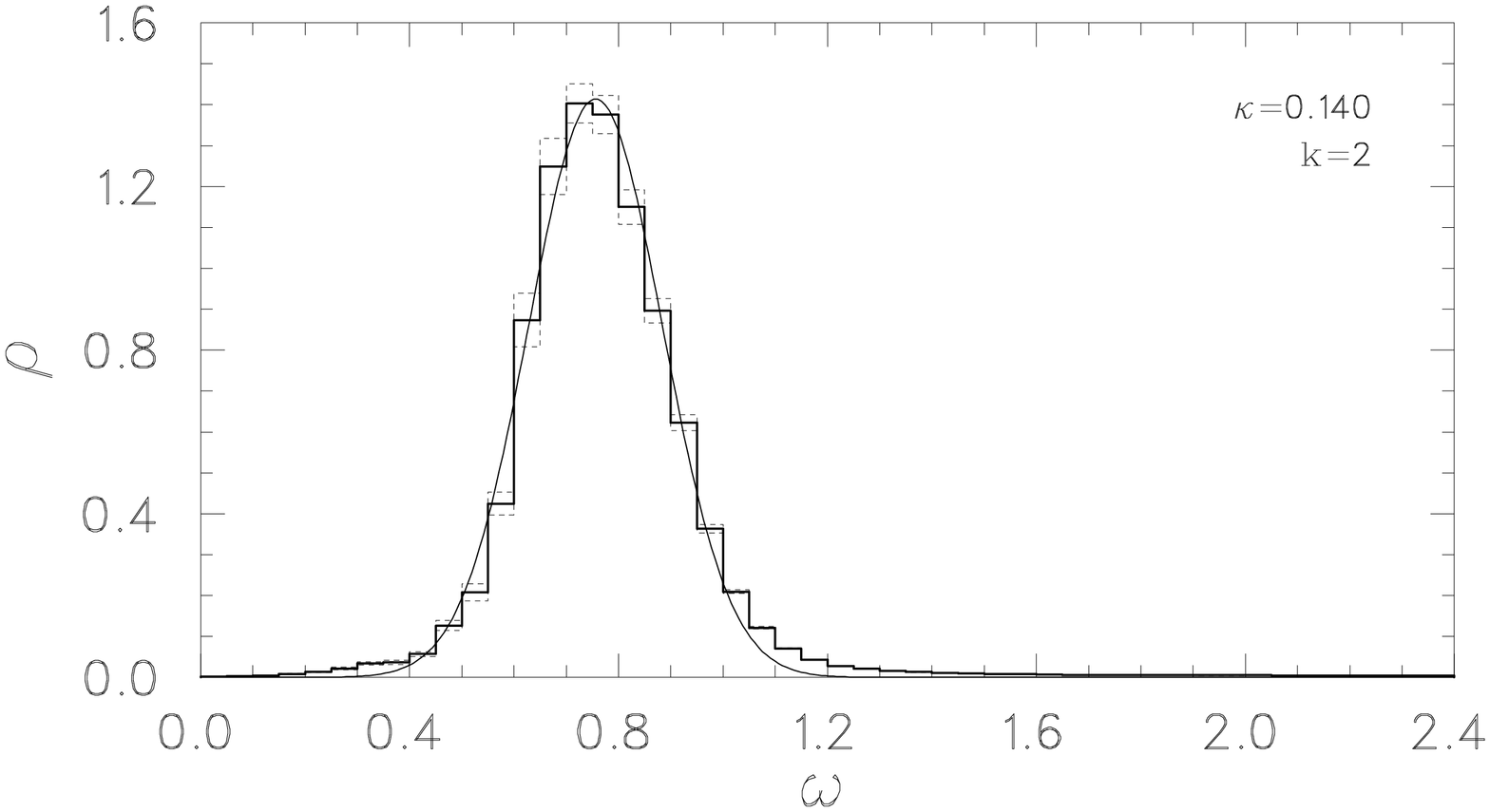}
\includegraphics[height=70mm,angle=90]{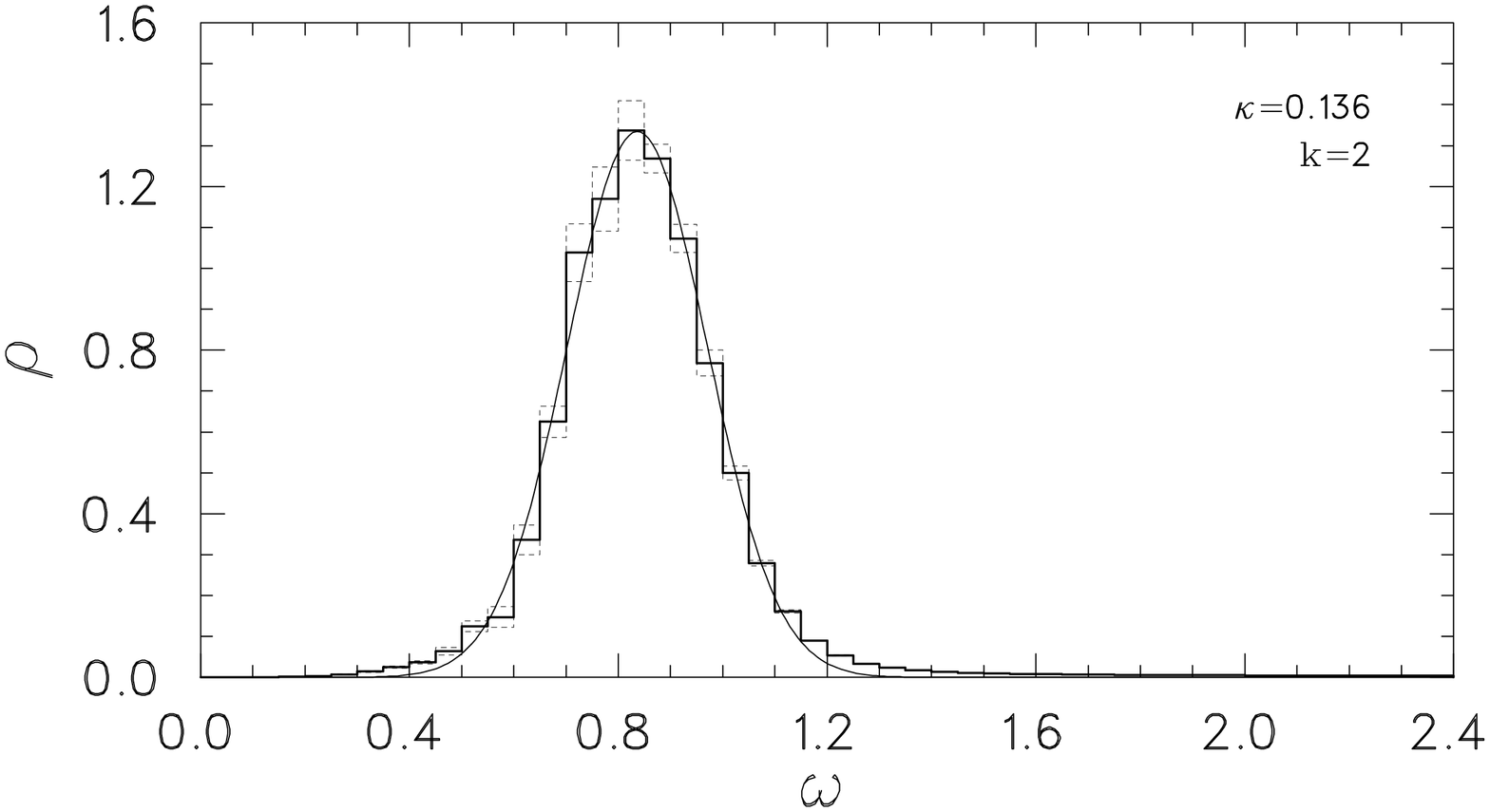}
\includegraphics[height=70mm,angle=90]{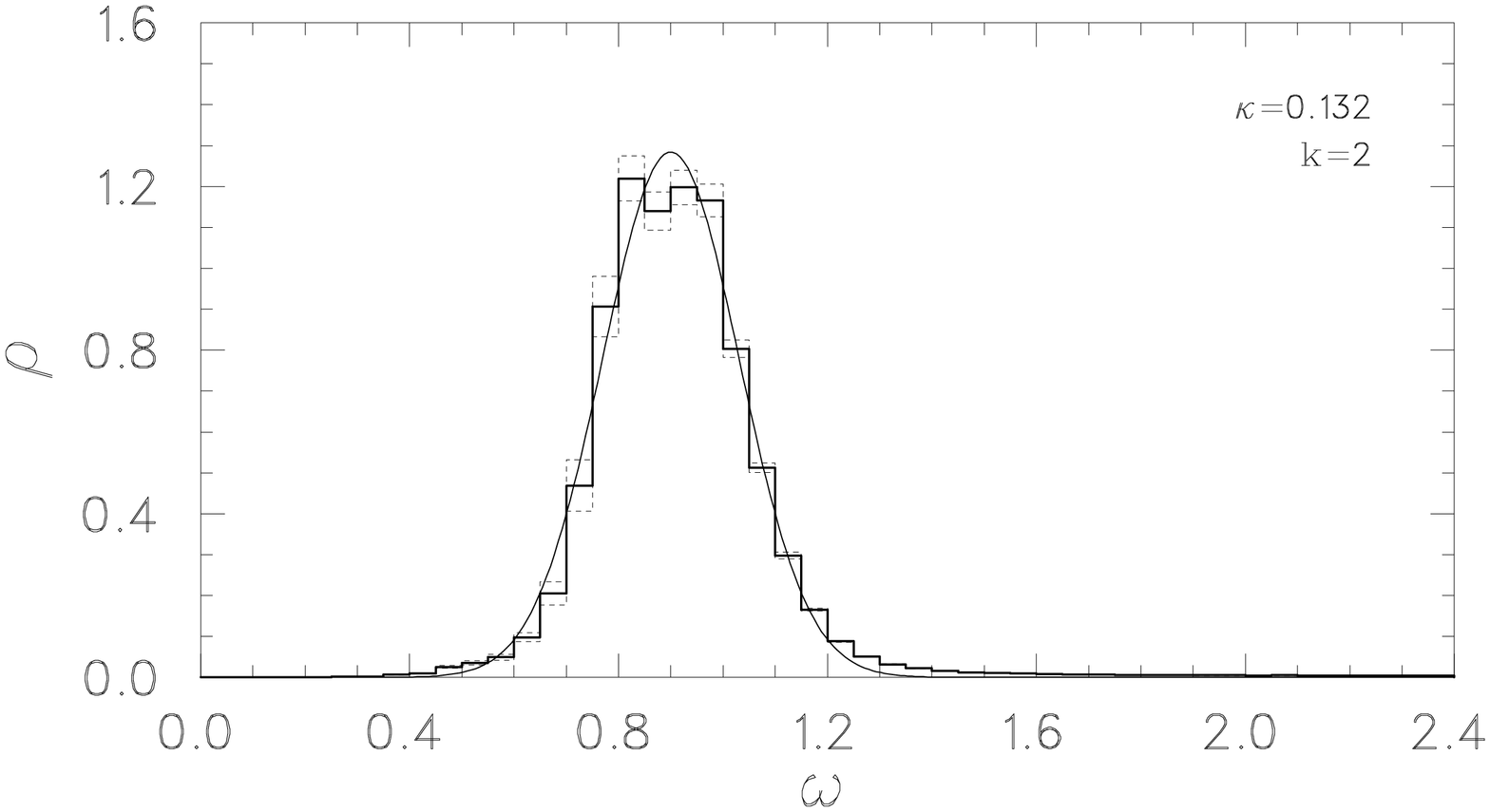}
\includegraphics[height=70mm,angle=90]{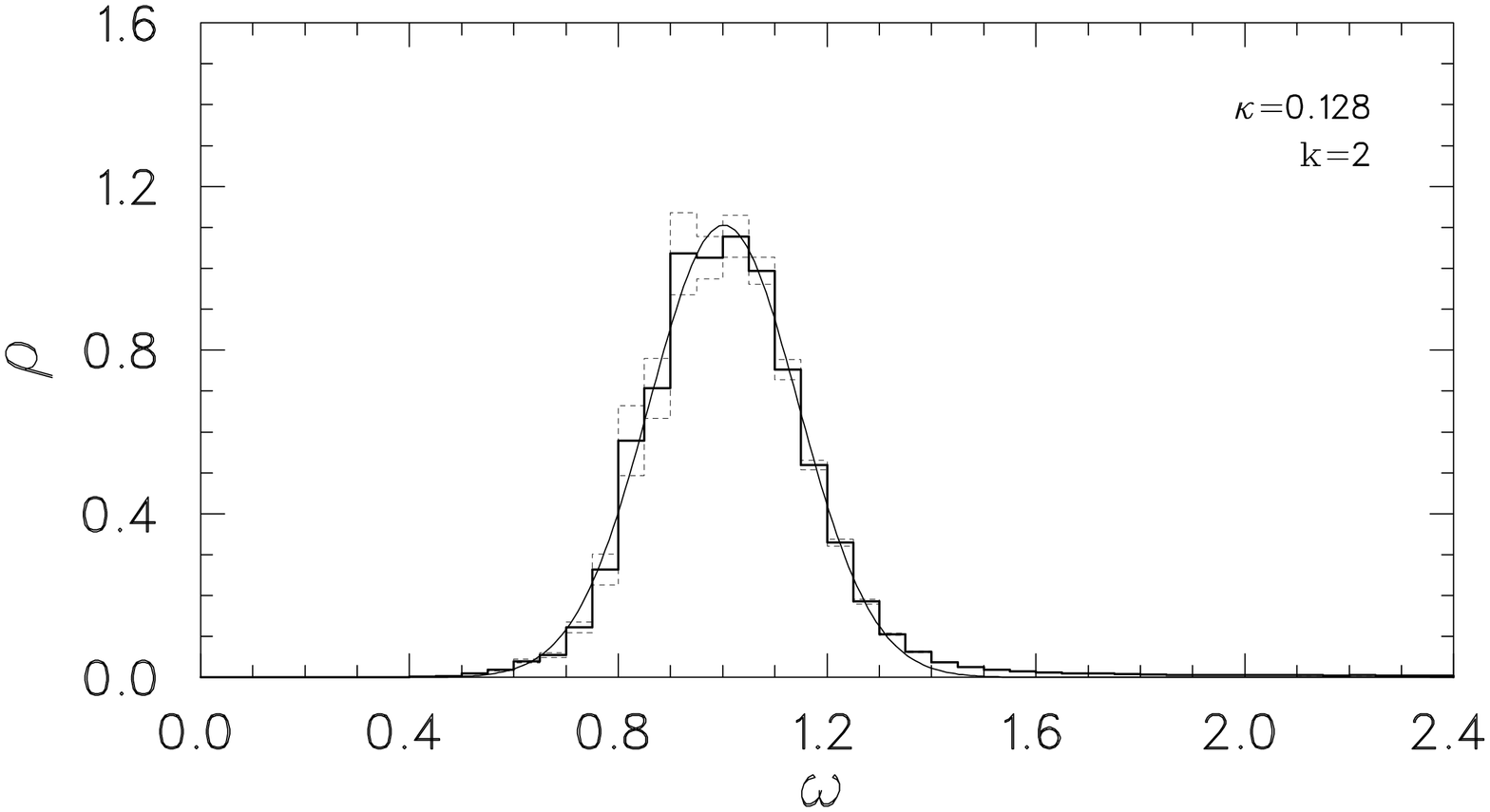}
\caption{\label{fig:memspecs2}As Fig.~\protect\ref{fig:memspecs1}, but for the
second (intermediate) eigenvalue set $\sigma_{\h,k}$, $k=2$.}
\end{figure}

\begin{figure}[t]
\includegraphics[height=71.5mm,angle=90]{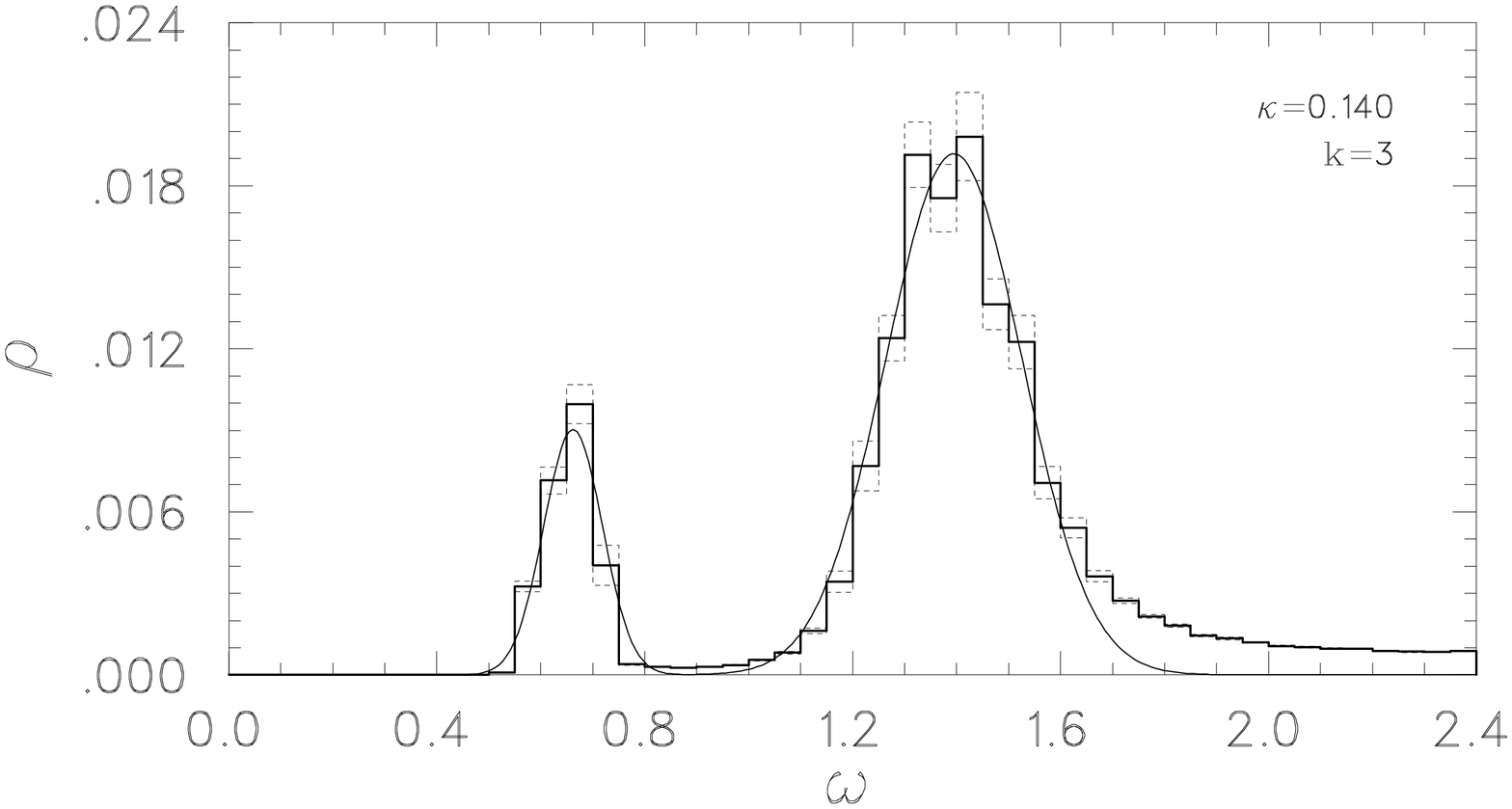}
\includegraphics[height=71.5mm,angle=90]{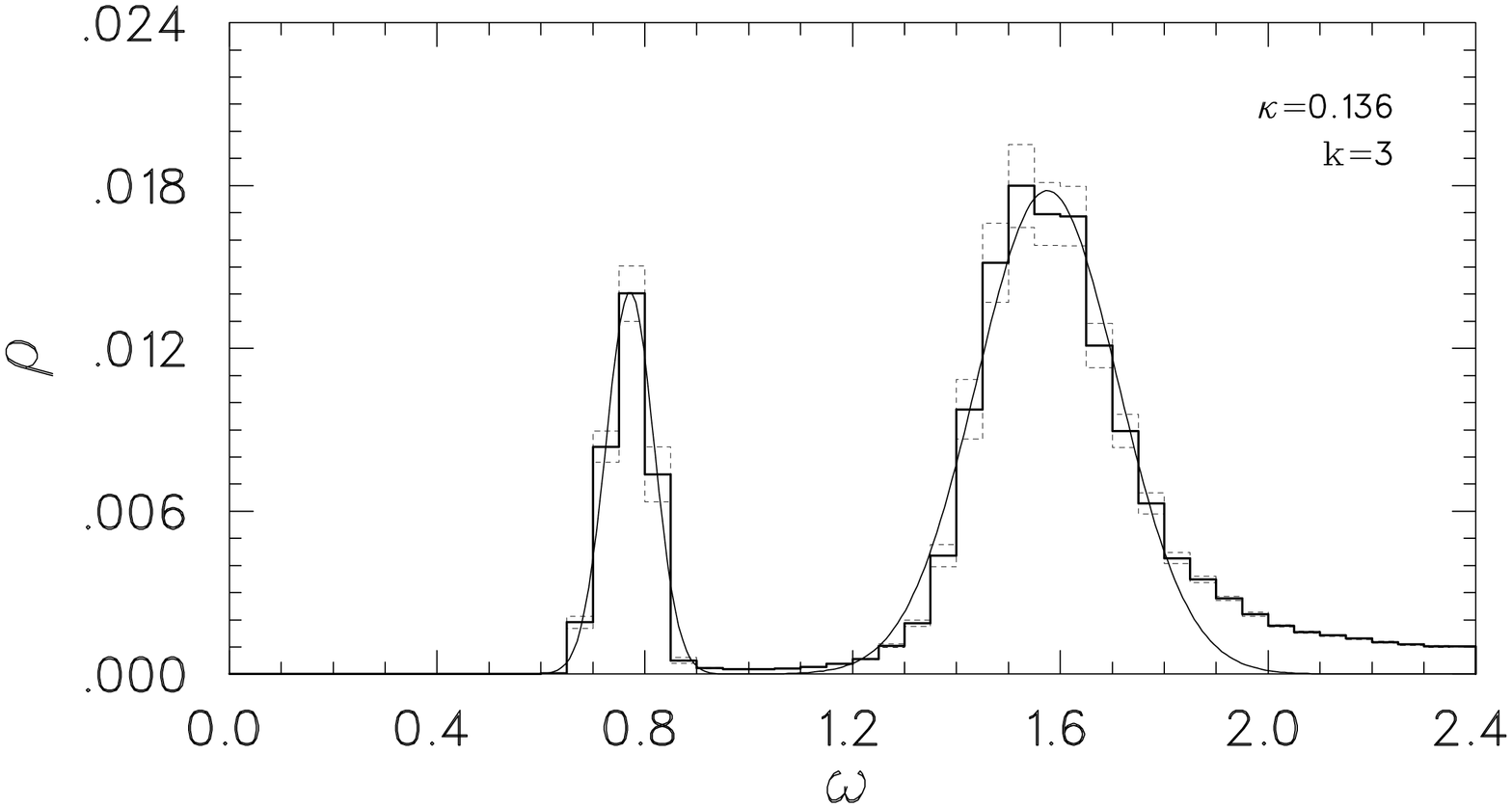}
\includegraphics[height=71.5mm,angle=90]{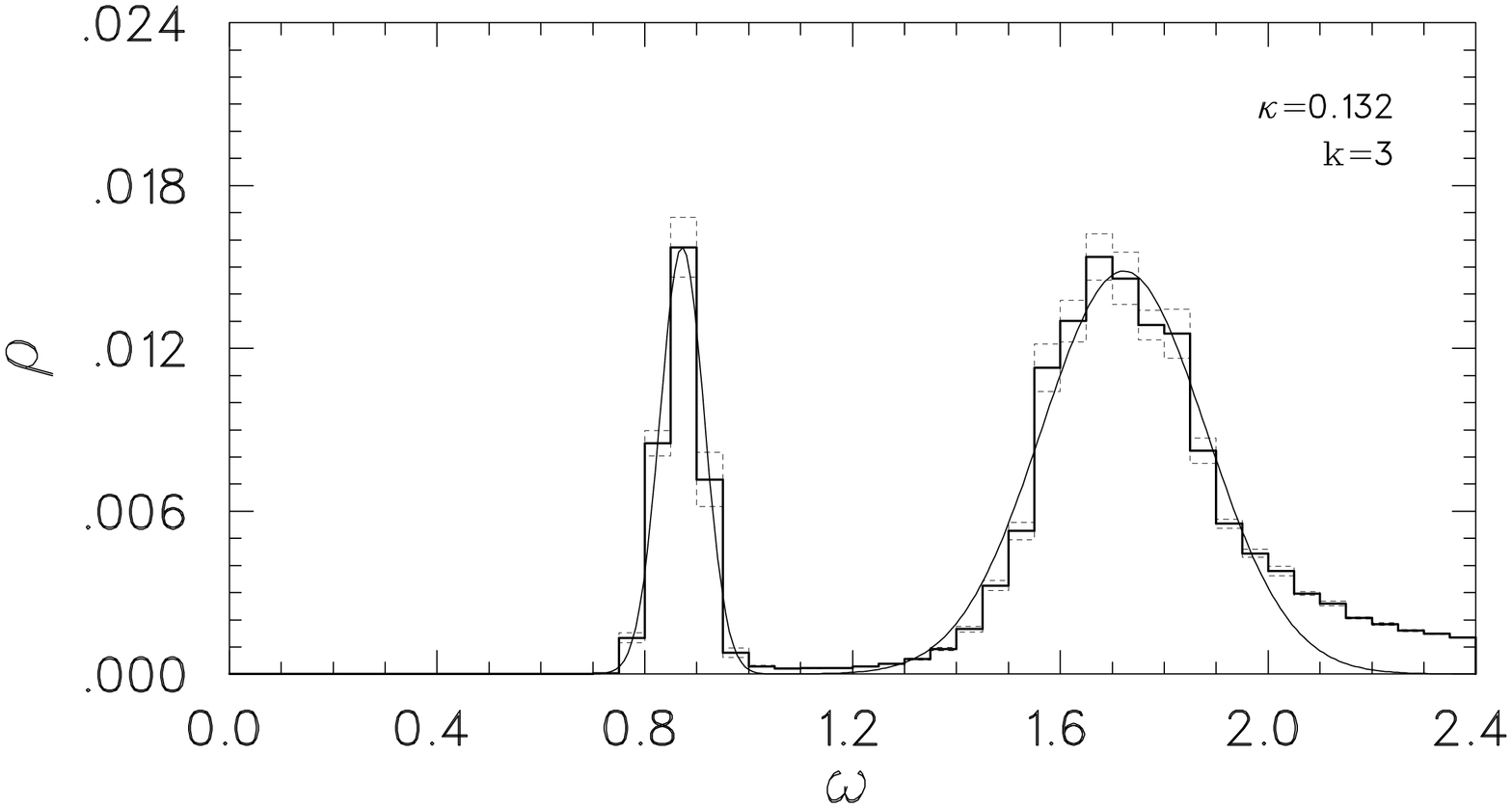}
\includegraphics[height=71.5mm,angle=90]{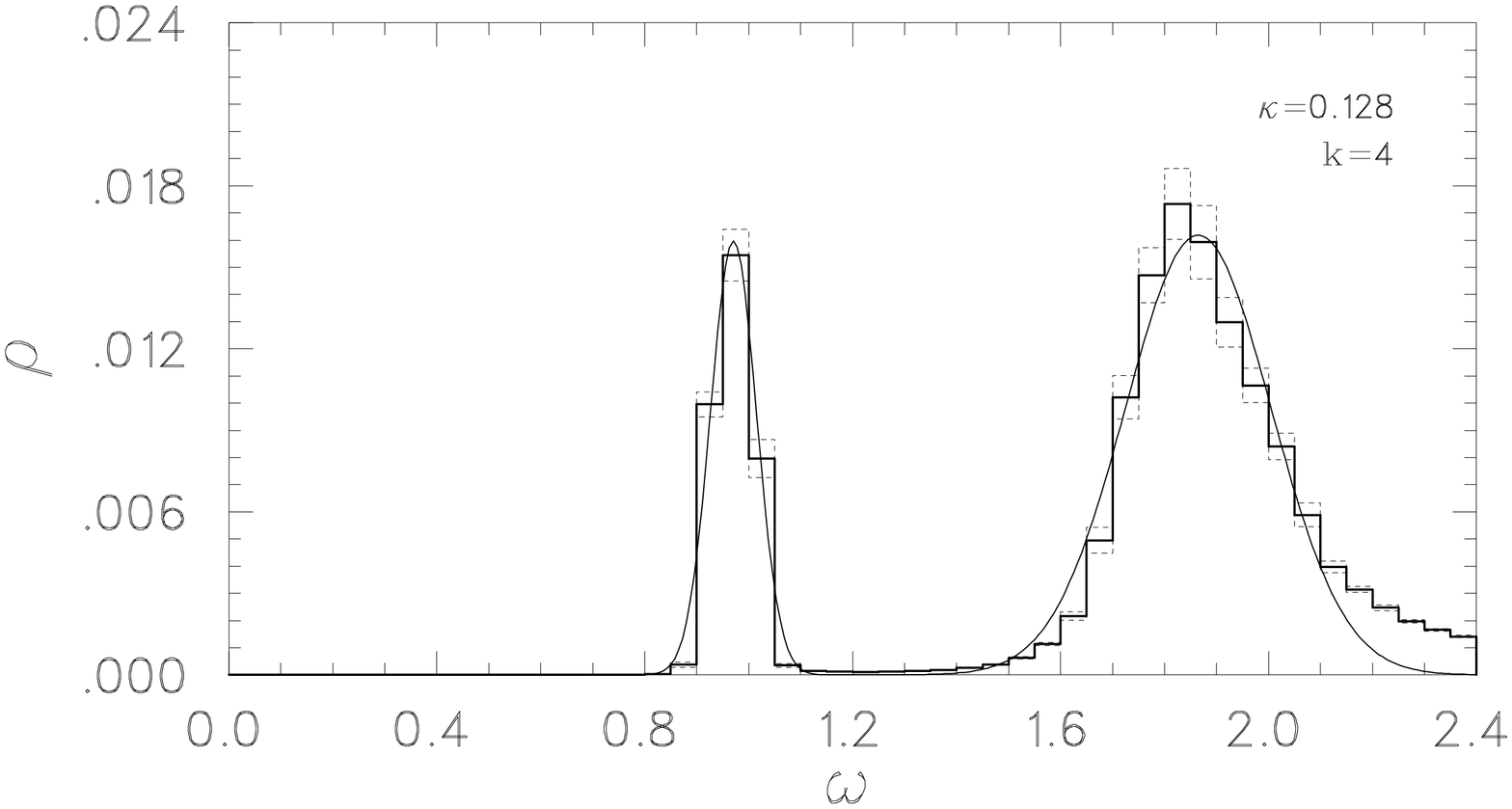}
\caption{\label{fig:memspecs3}As Fig.~\protect\ref{fig:memspecs1}, but for the
third (smallest) eigenvalue set $\sigma_{\h,k}$, $k=3$.}
\end{figure}

The spectral density functions $\rho(\omega)$ for the
eigenvalues $\sigma_{\h;k}$, $k=1,2,3$, of the hybrid
meson correlation matrix are displayed in
Figs.~\ref{fig:memspecs1},\ref{fig:memspecs2},\ref{fig:memspecs3} respectively.
The four panels in each of the
figures show the results for the four hopping parameter values as listed
in Tab.~\ref{tab:results} below.
The top panel, in each case, belongs to the largest $\kappa$, or the smallest pion mass,
and give rise to the model fits shown as solid lines in Fig.~\ref{fig:corrfx}.
Each spectral density comes from 16 random starts of the annealing process (\ref{Zmem}).
The center histogram (thick solid line) represents the average while the envelope histograms
(thin dashed line) indicate the standard deviation with respect to the cooling starts. 
   
In most cases the spectra exhibit isolated peaks, say 
$\delta_n=\{\omega|\omega\in{\rm peak}\ \#n\}$.
Then one may compute, for each peak, $n=1,2$, the volume $Z_n$,
the mass $E_n$, and the width $\Delta_n$, according to
\begin{eqnarray}
Z_n&=&\int_{\delta_n}d\omega\/\rho(\omega)\label{Zn}\\
E_n&=&Z_n^{-1}\int_{\delta_n}d\omega\/\rho(\omega)\omega\label{En}\\
\Delta_n^2&=&Z_n^{-1}\int_{\delta_n}d\omega\/\rho(\omega)\left(\omega-E_n\right)^2\,.
\label{Dn}\end{eqnarray}
Those (average) quantities constitute the information that can reasonably
be expected to result from the Bayesian analysis.

If overlapping peaks are present the above analysis gives inaccurate results.
However, one may then resolve
the spectral peaks by fitting a superposition of Gaussians to $\rho(\omega)$.
Because we are forced to do this in at least one case, see Fig.~\ref{fig:memspecs1},
we have consistently made fits with one or two Gaussians, as required, of
all spectral densities for the hybrid meson.
To do this a Levenberg-Marquardt algorithm was employed, and the initial set of
parameters were conveniently taken from applying (\ref{Zn})--(\ref{En}).
The results of those fits are shown as smooth solid lines in
Figs.~\ref{fig:memspecs1},\ref{fig:memspecs2},\ref{fig:memspecs3}.

A complete list of the results is given in Tab.~\ref{tab:results}. There, we also show the
MEM results for standard mesons $X=\pi,\rho,a_1$, which were obtained in the same manner,
i.e. from $3\times 3$ correlation matrices, etc. The masses of those mesons are used for
reference to the physical scale, and extrapolation to the physical pion mass region,
as discussed in the next section.
The peak volumes $Z_n$ are related to matrix elements $\langle n|\Phi_k(t)|0\rangle$,
where $\Phi_k(t)$ is a linear combination of the (three) operators used to
construct the correlator matrix \cite{Fiebig:2002sp}. As such they are a measure for how
strongly the set of operators excite states $|n\rangle$ from the vacuum. The overall
normalization, however, is arbitrary.
\begin{table}[t]
\caption{\label{tab:results} Results of the MEM analysis of correlator matrix eigenvalues
for meson operators $X$ at four values of the hopping parameter $\kappa$.
The number $k$ is the eigenvalue label.
The entries for the peak volume $Z_n$, the peak energy $E_n$, and the peak width $\Delta_n$
for the primary peak $n=1$ and secondary peak $n=2$, where appropriate, are the results
of Gaussian fits to the computed spectral density functions,
see Figs.~\protect\ref{fig:memspecs1}--\protect\ref{fig:memspecs3}.
The uncertainties are standard deviations resulting from 16 independent cooling starts.
The presence of $\pm 3$ indicates factors of $10^{\pm 3}$.}
\begin{ruledtabular}
\begin{tabular}{cccclll}
$X$ & ${\kappa}$ & $k$ & $n$ & $Z_n$ & $a_t E_n$ & $a_t \Delta_n$ \\
\colrule
$\pi$  & 0.140 & 1 & 1 & 18.4(0.8) & 0.50(1) & 0.0577(1) \\
       & 0.136 & 1 & 1 & 17.0(1.4) & 0.60(1) & 0.0577(6) \\
       & 0.132 & 1 & 1 & 16.1(1.4) & 0.71(1) & 0.057(1) \\
       & 0.128 & 1 & 1 & 14.9(1.2) & 0.82(1) & 0.055(3) \\
\colrule
$\rho$ & 0.140 & 1 & 1 & 46.0(1.9) & 0.535(7) & 0.072(8) \\
       & 0.136 & 1 & 1 & 47.1(0.7) & 0.628(8) & 0.067(6) \\
       & 0.132 & 1 & 1 & 44.4(1.3) & 0.725(7) & 0.070(8) \\
       & 0.128 & 1 & 1 & 42.6(0.5) & 0.828(4) & 0.067(5) \\
\colrule
$a_1$  & 0.140 & 1 & 1 & 30.1(1.0) & 0.69(1)$\circ$ & 0.13(1) \\
       & 0.136 & 1 & 1 & 21.7(0.9) & 0.77(1)$\circ$ & 0.11(1) \\
       & 0.132 & 1 & 1 & 17.2(0.4) & 0.875(9)$\circ$ & 0.105(8) \\
       & 0.128 & 1 & 1 & 12.8(0.3) & 0.979(9)$\circ$ & 0.096(9) \\
\colrule
$\h$   & 0.140 & 1 & 1 & 0.10(6)$+3$ & 0.62(4) & 0.16(1) \\
       & 0.136 & 1 & 1 & 0.08(2)$+3$ & 0.66(3) & 0.127(9) \\
       & 0.132 & 1 & 1 & 0.07(1)$+3$ & 0.76(2) & 0.079(4) \\
       & 0.128 & 1 & 1 & 0.09(1)$+3$ & 0.86(2) & 0.083(4) \\
$\h$   & 0.140 & 1 & 2 & 0.10(5)$+3$ & 0.95(9) & 0.3(2) \\
       & 0.136 & 1 & 2 & 0.123(9)$+3$ & 1.02(2) & 0.17(2) \\
       & 0.132 & 1 & 2 & 0.106(8)$+3$ & 1.13(2) & 0.17(3) \\
       & 0.128 & 1 & 2 & 0.079(6)$+3$ & 1.26(2) & 0.21(3) \\
\colrule
$\h$   & 0.140 & 2 & 1 & 0.455(8) & 0.756(7) & 0.26(2) \\
       & 0.136 & 2 & 1 & 0.449(9) & 0.837(9) & 0.27(2) \\
       & 0.132 & 2 & 1 & 0.421(5) & 0.900(6) & 0.26(1) \\
       & 0.128 & 2 & 1 & 0.395(8) & 1.002(8) & 0.29(2) \\
\colrule
$\h$   & 0.140 & 3 & 1 & 1.28(8)$-3$ & 0.662(8)$\times$ & 0.11(3) \\
       & 0.136 & 3 & 1 & 1.6(1)$-3$ & 0.771(9)$\times$ & 0.09(3) \\
       & 0.132 & 3 & 1 & 1.7(2)$-3$ & 0.872(8)$\times$ & 0.09(3) \\
       & 0.128 & 3 & 1 & 1.77(8)$-3$ & 0.971(5)$\times$ & 0.09(2) \\
$\h$   & 0.140 & 3 & 2 & 6.3(2)$-3$ & 1.395(8) & 0.26(4) \\
       & 0.136 & 3 & 2 & 6.1(2)$-3$ & 1.58(1) & 0.27(3) \\
       & 0.132 & 3 & 2 & 5.9(1)$-3$ & 1.72(1) & 0.32(2) \\
       & 0.128 & 3 & 2 & 5.7(2)$-3$ & 1.87(1) & 0.28(4)
\end{tabular}
\end{ruledtabular}
\end{table}

Some comments on uncertainties are in order: As a matter of course jackknife
configurations \cite{Efr79} were generated to estimate statistical errors.
Those are reflected in the error bars of the eigenvalue correlator plots of
Fig.~\ref{fig:corrfx}.
However, in the ensuing MEM analysis jackknife errors are hard to maintain,
because extraction of the spectral density by way of simulated annealing
is very time intensive. (Ideally, the cooling steps should be infinitely small.)
For this reason the MEM analysis was only applied to gauge configuration
averages. Nevertheless, the peak widths of the spectral density functions
naturally emerge as a measure of uncertainties for the masses.
One may even argue that those are the `true' measures of uncertainties
if compared to statistical (jackknife) errors, because the latter can in
principle be made arbitrarily small by increasing the number of gauge
configurations, even though the lattice data may be poor.
Peak width uncertainties and plateau statistical errors are hard to relate,
one reason being that they stem from different data sets.
This notwithstanding we observe that the masses of Ref.~\cite{Cook:2006tz},
which carry statistical errors, are generally consistent with the  
ground state meson masses obtained in the current analysis using peak
width uncertainties.

\section{\label{sec:dis}Discussion of results}

\begin{figure}[h]
\includegraphics[width=86mm,angle=0]{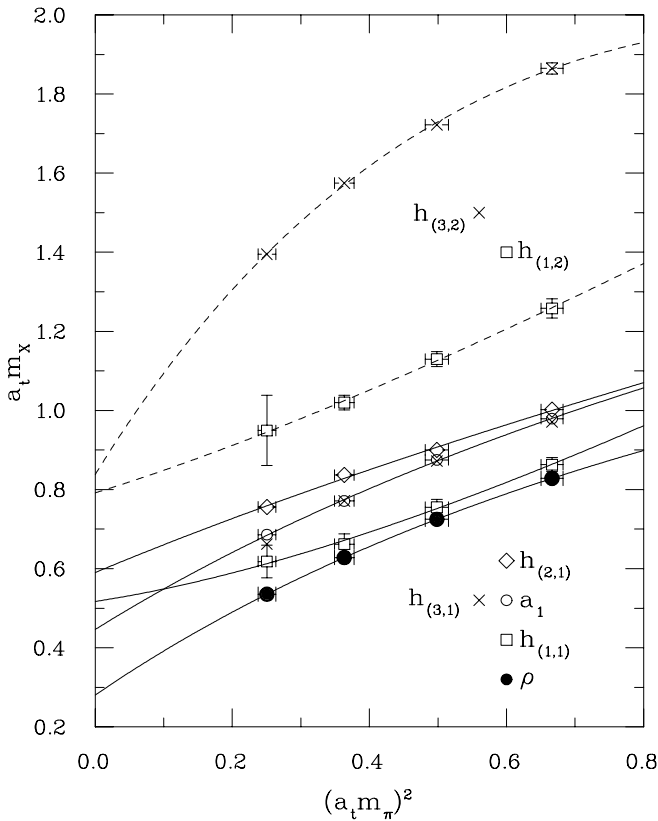}
\caption{\label{fig:rscale}Plots of spectral meson masses
versus the squared pion mass $x=(a_tm_\pi)^2$ and fits with the model
(\protect\ref{fitln}). Solid lines refer to masses from primary ($n=1$) spectral peaks,
and dashed lines to secondary ($n=2$) peaks, see Tab.~\protect\ref{tab:results}. 
The extrapolation of the $\rho$ meson mass to $x=0$ is used to set
the physical scale.}
\end{figure}

In order to set the spectral masses into a physical context we have attempted
extrapolations to the physical pion mass region. Thus we consider the computed
$\rho,a_{1}$ and $\h$ spectral masses as functions of $m_\pi^2$.
To our knowledge, predictions for this dependence from chiral perturbation
\cite{BernsteinHolstein} are not available, as a hybrid meson is involved.
We therefore employ a heuristic model. It has been used in Ref.~\cite{Cook:2006tz},
where also some motivation was given.
For a spectral mass $M=a_{t}m$ we attempt fits with the model
\begin{equation}\label{fitln}
M=p+qx+r\ln (1+x) \quad\,\, \mbox{with} \,\,\, x=(a_{t}m_\pi)^2\,,
\end{equation}
using three parameters $p,q,r$. The linear term is standard in $\chi$PT inspired models
while the logarithmic term is purely heuristical but happens to yield excellent fits
to our data.
The results are illustrated in Fig.~\ref{fig:rscale}.

The extrapolated $\rho$ meson mass, at $x=0$, is used to determine the physical scale
for this simulation. We obtain 
$a_t=0.36(6)\,\mbox{GeV}^{-1}\,=0.072(11)\,\mbox{fm}$ ($m_{\rho}=775.8\,\mbox{MeV}$).
Table~\ref{tab:extrapol} then gives a list of the extrapolated masses $a_tE_X$
along with the physical values $E_X$. Also listed are estimates of the peak widths at $x=0$.
Those are obtained by randomizing the data,
replacing the four data points with $a_tE_n+\xi a_t\Delta_n$
where $\xi$ is a normal distributed random deviate with variance one,
and then repeating the fits with the model (\ref{fitln}) obtaining random values for the
extrapolated spectral masses $a_tE_X$.
A few thousand randomizations then give the results $\Delta_X$ in Tab.~\ref{tab:extrapol}.
Uncertainties on $\Delta_X$, being second order, were not calculated.
The errors given in parentheses are obtained in the same way, but using the uncertainties
from Tab.~\ref{tab:results} instead of the $\Delta_n$.
Unfortunately, extrapolation tends to amplify the peak widths because they become larger
at smaller pion masses. As a measure of uncertainty the extrapolated $\Delta_X$,
in Tab.~\ref{tab:extrapol}, thus seem less useful.
\begin{table}[h]
\caption{\label{tab:extrapol} Extrapolated spectral masses $E_X$ and peak widths $\Delta_X$,
for mesons $X$. The eigenvalue label is $k$ and the spectral peak number is $n$,
as in Tab.~\protect\ref{tab:results}.
The values for $\Delta_X$ and for the uncertainties (in parentheses) are obtained by randomization
of the data points, as explained in the text.}
\begin{ruledtabular}
\begin{tabular}{ccclclc}
$X$ & $k$ & $n$ & $a_t E_X$ & $a_t \Delta_X$ & $E_X$[GeV] & $\Delta_X$[GeV] \\
\colrule
$\rho$ & 1 & 1 & 0.28(04) & 0.08 & 0.7785     & 0.22 \\
\colrule
$a_1$  & 1 & 1 & 0.45(06) & 0.17 & 1.23(0.17) & 0.47 \\
\colrule
$\h$   & 1 & 1 & 0.52(19) & 0.37 & 1.43(0.53) & 1.03 \\
$\h$   & 1 & 2 & 0.79(37) & 0.65 & 2.19(1.03) & 1.80 \\
\colrule
$\h$   & 2 & 1 & 0.59(04) & 0.27 & 1.63(0.12) & 0.76 \\
\colrule
$\h$   & 3 & 1 & 0.34(05)$\times$ & 0.20 & 0.94(0.13)$\times$ & 0.55 \\
$\h$   & 3 & 2 & 0.84(07) & 0.11 & 2.32(0.18) & 0.30
\end{tabular}
\end{ruledtabular}
\end{table}

As a first observation we note that the $a_1$ meson mass in Tab.~\ref{tab:extrapol} comes
out very close to the experimental value $1230$ MeV, which is the mass of
the $a_1(1260)$ meson, in the nomenclature of \cite{PDBook:2004}.
We take this as a hint that the extrapolation model is adequate,
and therefore, the extrapolated results for the $\h$ mass spectrum,
in Tab.~\ref{tab:extrapol}, may not be without merit.

In the light of this we continue to observe that the energy levels
of $X_{(k,n)}$ for $\h_{(1,1)}$
and $\h_{(2,1)}$ in Tab.~\ref{tab:extrapol} match the experimental masses 1376 MeV and
1653 MeV of the $1^{-+}$ resonances known as $\pi_1(1400)$
and $\pi_1(1600)$ in Ref.~\cite{PDBook:2004}.
One should not overinterpret the closeness of the computed masses to the experimental levels, 
because systematic errors associated with the extrapolation could be large.
On the other hand the spectral level pattern of two $1^{-+}$ states below 2 GeV is
a unequivocal result of this simulation, which in no small part relies on the MEM analysis of the
correlator data.
It is also interesting to note that the extrapolated spectral level sequence only
emerges beyond the level crossing of the states $\h_{(1,1)}$ and $a_1$ at around
$x\simeq 0.1$, see Fig.~\ref{fig:rscale}.

Another remarkable results concerns the level $\h_{(3,1)}$. Its
energy spectrum coincides with the one generated by the $a_1$ meson operator.   
In Tab.~\ref{tab:results} the symbols $\circ$ and $\times$ are attached to the
corresponding levels. These are also used in Fig.~\ref{fig:rscale}, where an almost
perfect match is apparent, only the data point at the smallest pion mass is slightly off.
It causes, however, unrealistic extrapolation results, see the entries in
Tab.~\ref{tab:extrapol} marked with a $\times$.
The key to an explanation of this curious result goes back to the observation made in
Sect.~\ref{sec:lat}, namely that parity is guaranteed only in the limit of a large
number of gauge configurations. The level $\h_{(3,1)}$ comes from a correlator eigenvalue
about four orders of magnitude less than the dominant one. It may thus single out
the contamination of the lattice signal with a $1^{++}$ state, which happen to be
the quantum numbers of the $a_1$ meson \cite{PDBook:2004}.
Although this observation is not directly relevant to our current interest in $1^{-+}$
exotic states, it nevertheless adds credence to the current simulation and analysis method.

Lastly, the largest levels $\h_{(1,2)}$ and $\h_{(3,2)}$ in Tab.~\ref{tab:extrapol}
point at a mass somewhat above 2 GeV.
We speculate that at least one of those levels coincides with the $1^{-+}$ resonance
at 1.9 GeV uncovered in Ref.~\cite{Cook:2006tz}. There the space of operators was larger,
including a $\pi a_1$ two-hadron field in addition to $h$, and this could be the cause
for a lowering of the energy level from 2 GeV to 1.9 GeV. If this interpretation is correct, 
then this state likely is a mixture of a hybrid meson and a two-meson state.
We take this result as supplementary evidence to the outcome of Ref.~\cite{Cook:2006tz}.

\section{Conclusion}

Using the maximum entropy method, five distinct spectral levels have been 
uncovered for the $J^{PC}=1^{-+}$ exotic meson. Two of the spectral levels
correspond with the $\pi_1(1400)$ and $\pi_1(1600)$ from \cite{PDBook:2004}.
Two more levels possibly correspond with a resonance energy of $1.9$ GeV 
previously determined by a decay width calculation using L{\"u}scher's method
\cite{Cook:2006tz}. A fifth spectral level, at higher pion masses, tracks consistent
with an operator representing the $a_1(1260)$ meson, and we take this level to
be a consequence of inexact parity symmetry.

All of these spectral levels rely on extrapolations to $m_{\pi}=0$ from relatively
heavy pions. Although this may give rise to large systematic errors, the fact 
that the $a_1$ extrapolation came very close to its experimental value leads 
us to conclude at least two spectral levels for the $1^{-+}$ 
exotic meson will be below $2$ GeV.


\begin{thebibliography}{24}
\expandafter\ifx\csname natexlab\endcsname\relax\def\natexlab#1{#1}\fi
\expandafter\ifx\csname bibnamefont\endcsname\relax
  \def\bibnamefont#1{#1}\fi
\expandafter\ifx\csname bibfnamefont\endcsname\relax
  \def\bibfnamefont#1{#1}\fi
\expandafter\ifx\csname citenamefont\endcsname\relax
  \def\citenamefont#1{#1}\fi
\expandafter\ifx\csname url\endcsname\relax
  \def\url#1{\texttt{#1}}\fi
\expandafter\ifx\csname urlprefix\endcsname\relax\def\urlprefix{URL }\fi
\providecommand{\bibinfo}[2]{#2}
\providecommand{\eprint}[2][]{\url{#2}}

\bibitem[{\citenamefont{Barnes}(2003)}]{Barnes:2003vy}
\bibinfo{author}{\bibfnamefont{T.}~\bibnamefont{Barnes}}
  (\bibinfo{year}{2003}), \eprint{nucl-th/0303032}.

\bibitem[{\citenamefont{Eidelman et~al.}(2004)}]{PDBook:2004}
\bibinfo{author}{\bibfnamefont{S.}~\bibnamefont{Eidelman}}
  \bibnamefont{et~al.}, \bibinfo{journal}{Physics Letters B}
  \textbf{\bibinfo{volume}{592}}, \bibinfo{pages}{1+} (\bibinfo{year}{2004}),
  \urlprefix\url{http://pdg.lbl.gov}.

\bibitem[{\citenamefont{Michael}(2005)}]{Michael:2005tw}
\bibinfo{author}{\bibfnamefont{C.}~\bibnamefont{Michael}}, in
  \emph{\bibinfo{booktitle}{23rd International Symposium on Lattice Field
  Theory, 25-30 July 2005, Trinity College, Dublin, Ireland, PoS(LAT2005)008}}
  (\bibinfo{year}{2005}), \eprint{hep-lat/0509023}.

\bibitem[{\citenamefont{Hedditch et~al.}(2005)}]{Hedditch:2005zf}
\bibinfo{author}{\bibfnamefont{J.~N.} \bibnamefont{Hedditch}}
  \bibnamefont{et~al.}, \bibinfo{journal}{Phys. Rev.}
  \textbf{\bibinfo{volume}{D72}}, \bibinfo{pages}{114507}
  (\bibinfo{year}{2005}), \eprint{hep-lat/0509106}.

\bibitem[{\citenamefont{Bernard et~al.}(1997)}]{Bernard:1997ib}
\bibinfo{author}{\bibfnamefont{C.~W.} \bibnamefont{Bernard}}
  \bibnamefont{et~al.} (\bibinfo{collaboration}{MILC}), \bibinfo{journal}{Phys.
  Rev.} \textbf{\bibinfo{volume}{D56}}, \bibinfo{pages}{7039}
  (\bibinfo{year}{1997}), \eprint[http://arXiv.org/abs]{hep-lat/9707008}.

\bibitem[{\citenamefont{McNeile}(2002)}]{McNeile:2002en}
\bibinfo{author}{\bibfnamefont{C.}~\bibnamefont{McNeile}},
  \bibinfo{journal}{Nucl. Phys.} \textbf{\bibinfo{volume}{A711}},
  \bibinfo{pages}{303} (\bibinfo{year}{2002}), \eprint{hep-lat/0207001}.

\bibitem[{\citenamefont{Cook and Fiebig}(2006)}]{Cook:2006tz}
\bibinfo{author}{\bibfnamefont{M.~S.} \bibnamefont{Cook}} \bibnamefont{and}
  \bibinfo{author}{\bibfnamefont{H.~R.} \bibnamefont{Fiebig}},
  \bibinfo{journal}{Phys. Rev.} \textbf{\bibinfo{volume}{D74}},
  \bibinfo{pages}{034509} (\bibinfo{year}{2006}), \eprint{hep-lat/0606005}.

\bibitem[{\citenamefont{McNeile and Michael}(2006)}]{McNeile2006}
\bibinfo{author}{\bibfnamefont{C.}~\bibnamefont{McNeile}} \bibnamefont{and}
  \bibinfo{author}{\bibfnamefont{C.}~\bibnamefont{Michael}}
  (\bibinfo{year}{2006}), \eprint{hep-lat/0603007}.

\bibitem[{\citenamefont{McNeile and Michael}(2003)}]{McNeile:2002fh}
\bibinfo{author}{\bibfnamefont{C.}~\bibnamefont{McNeile}} \bibnamefont{and}
  \bibinfo{author}{\bibfnamefont{C.}~\bibnamefont{Michael}}
  (\bibinfo{collaboration}{UKQCD}), \bibinfo{journal}{Phys. Lett.}
  \textbf{\bibinfo{volume}{B556}}, \bibinfo{pages}{177} (\bibinfo{year}{2003}),
  \eprint{hep-lat/0212020}.

\bibitem[{\citenamefont{McNeile et~al.}(2002)\citenamefont{McNeile, Michael,
  and Pennanen}}]{McNeile:2002az}
\bibinfo{author}{\bibfnamefont{C.}~\bibnamefont{McNeile}},
  \bibinfo{author}{\bibfnamefont{C.}~\bibnamefont{Michael}}, \bibnamefont{and}
  \bibinfo{author}{\bibfnamefont{P.}~\bibnamefont{Pennanen}}
  (\bibinfo{collaboration}{UKQCD}), \bibinfo{journal}{Phys. Rev.}
  \textbf{\bibinfo{volume}{D65}}, \bibinfo{pages}{094505}
  (\bibinfo{year}{2002}), \eprint[http://arXiv.org/abs]{hep-lat/0201006}.

\bibitem[{\citenamefont{Jarrell and Gubernatis}(1996)}]{Jar96}
\bibinfo{author}{\bibfnamefont{M.}~\bibnamefont{Jarrell}} \bibnamefont{and}
  \bibinfo{author}{\bibfnamefont{J.~E.} \bibnamefont{Gubernatis}},
  \bibinfo{journal}{Phys. Rep.} \textbf{\bibinfo{volume}{269}},
  \bibinfo{pages}{133} (\bibinfo{year}{1996}).

\bibitem[{\citenamefont{Gupta}(1998)}]{Gupta:1998tz}
\bibinfo{author}{\bibfnamefont{R.}~\bibnamefont{Gupta}}, in
  \emph{\bibinfo{booktitle}{Lectures given at the LXVIII Les Houches Summer
  School: Probing the Standard Model of Particle Interactions}}
  (\bibinfo{year}{1998}), \eprint{hep-lat/9807028}.

\bibitem[{\citenamefont{Alexandrou et~al.}(1994)\citenamefont{Alexandrou,
  G{\"u}sken, Jegerlehner, Schilling, and Sommer}}]{Alexandrou:1994ti}
\bibinfo{author}{\bibfnamefont{C.}~\bibnamefont{Alexandrou}},
  \bibinfo{author}{\bibfnamefont{S.}~\bibnamefont{G{\"u}sken}},
  \bibinfo{author}{\bibfnamefont{F.}~\bibnamefont{Jegerlehner}},
  \bibinfo{author}{\bibfnamefont{K.}~\bibnamefont{Schilling}},
  \bibnamefont{and} \bibinfo{author}{\bibfnamefont{R.}~\bibnamefont{Sommer}},
  \bibinfo{journal}{Nucl. Phys.} \textbf{\bibinfo{volume}{B414}},
  \bibinfo{pages}{815} (\bibinfo{year}{1994}),
  \eprint[http://arXiv.org/abs]{hep-lat/9211042}.

\bibitem[{\citenamefont{Albanese et~al.}(1987)}]{Alb87a}
\bibinfo{author}{\bibfnamefont{C.}~\bibnamefont{Albanese}}
  \bibnamefont{et~al.}, \bibinfo{journal}{Phys.\ Lett.}
  \textbf{\bibinfo{volume}{B192}}, \bibinfo{pages}{163} (\bibinfo{year}{1987}).

\bibitem[{\citenamefont{L{\"u}scher and Wolff}(1990)}]{Luscher:1990ck}
\bibinfo{author}{\bibfnamefont{M.}~\bibnamefont{L{\"u}scher}} \bibnamefont{and}
  \bibinfo{author}{\bibfnamefont{U.}~\bibnamefont{Wolff}},
  \bibinfo{journal}{Nucl. Phys.} \textbf{\bibinfo{volume}{B339}},
  \bibinfo{pages}{222} (\bibinfo{year}{1990}).

\bibitem[{\citenamefont{Golub and Van~Loan}(1996)}]{Golub96}
\bibinfo{author}{\bibfnamefont{G.~H.} \bibnamefont{Golub}} \bibnamefont{and}
  \bibinfo{author}{\bibfnamefont{C.~F.} \bibnamefont{Van~Loan}},
  \emph{\bibinfo{title}{Matrix Computations}} (\bibinfo{publisher}{Johns
  Hopkins University Press}, \bibinfo{address}{Baltimore, MD},
  \bibinfo{year}{1996}), \bibinfo{edition}{3rd} ed., \bibinfo{note}{iSBN
  0-8018-5413-X}.

\bibitem[{Efr()}]{Efr79}
\bibinfo{note}{B. Efron, SIAM Review 21 (1979) 460}.

\bibitem[{\citenamefont{Yamazaki et~al.}(2002)}]{Yamazaki:2001er}
\bibinfo{author}{\bibfnamefont{T.}~\bibnamefont{Yamazaki}} \bibnamefont{et~al.}
  (\bibinfo{collaboration}{CP-PACS}), \bibinfo{journal}{Phys. Rev.}
  \textbf{\bibinfo{volume}{D65}}, \bibinfo{pages}{014501}
  (\bibinfo{year}{2002}), \eprint[http://arXiv.org/abs]{hep-lat/0105030}.

\bibitem[{\citenamefont{Asakawa et~al.}(2001)\citenamefont{Asakawa, Hatsuda,
  and Nakahara}}]{Asakawa:2000tr}
\bibinfo{author}{\bibfnamefont{M.}~\bibnamefont{Asakawa}},
  \bibinfo{author}{\bibfnamefont{T.}~\bibnamefont{Hatsuda}}, \bibnamefont{and}
  \bibinfo{author}{\bibfnamefont{Y.}~\bibnamefont{Nakahara}},
  \bibinfo{journal}{Prog. Part. Nucl. Phys.} \textbf{\bibinfo{volume}{46}},
  \bibinfo{pages}{459} (\bibinfo{year}{2001}),
  \eprint[http://arXiv.org/abs]{hep-lat/0011040}.

\bibitem[{\citenamefont{Asakawa et~al.}(2000)\citenamefont{Asakawa, Nakahara,
  and Hatsuda}}]{Asakawa:2000pv}
\bibinfo{author}{\bibfnamefont{M.}~\bibnamefont{Asakawa}},
  \bibinfo{author}{\bibfnamefont{Y.}~\bibnamefont{Nakahara}}, \bibnamefont{and}
  \bibinfo{author}{\bibfnamefont{T.}~\bibnamefont{Hatsuda}},
  \bibinfo{journal}{Nucl. Phys. Proc. Suppl.} \textbf{\bibinfo{volume}{86}},
  \bibinfo{pages}{191} (\bibinfo{year}{2000}).

\bibitem[{\citenamefont{Lepage et~al.}(2002)\citenamefont{Lepage, Clark,
  Davies, Hornbostel, Mackenzie, Morningstar, and Trottier}}]{Lepage:2001ym}
\bibinfo{author}{\bibfnamefont{G.~P.} \bibnamefont{Lepage}},
  \bibinfo{author}{\bibfnamefont{B.}~\bibnamefont{Clark}},
  \bibinfo{author}{\bibfnamefont{C.~T.~H.} \bibnamefont{Davies}},
  \bibinfo{author}{\bibfnamefont{K.}~\bibnamefont{Hornbostel}},
  \bibinfo{author}{\bibfnamefont{P.~B.} \bibnamefont{Mackenzie}},
  \bibinfo{author}{\bibfnamefont{C.}~\bibnamefont{Morningstar}},
  \bibnamefont{and} \bibinfo{author}{\bibfnamefont{H.}~\bibnamefont{Trottier}},
  \bibinfo{journal}{Nucl. Phys. Proc. Suppl.} \textbf{\bibinfo{volume}{106}},
  \bibinfo{pages}{12} (\bibinfo{year}{2002}),
  \eprint[http://arXiv.org/abs]{hep-lat/0110175}.

\bibitem[{\citenamefont{Fiebig}(2002{\natexlab{a}})}]{Fiebig:2001mr}
\bibinfo{author}{\bibfnamefont{H.~R.} \bibnamefont{Fiebig}}
  (\bibinfo{collaboration}{LHPC}), \bibinfo{journal}{Nucl. Phys. Proc. Suppl.}
  \textbf{\bibinfo{volume}{106}}, \bibinfo{pages}{344}
  (\bibinfo{year}{2002}{\natexlab{a}}),
  \eprint[http://arXiv.org/abs]{hep-lat/0110163}.

\bibitem[{\citenamefont{Fiebig}(2002{\natexlab{b}})}]{Fiebig:2002sp}
\bibinfo{author}{\bibfnamefont{H.~R.} \bibnamefont{Fiebig}},
  \bibinfo{journal}{Phys. Rev.} \textbf{\bibinfo{volume}{D65}},
  \bibinfo{pages}{094512} (\bibinfo{year}{2002}{\natexlab{b}}),
  \eprint[http://arXiv.org/abs]{hep-lat/0204004}.

\bibitem[{Ber(1995)}]{BernsteinHolstein}
in \emph{\bibinfo{booktitle}{Chiral Dynamics: Theory and Experiment}}, edited
  by \bibinfo{editor}{\bibfnamefont{A.~M.} \bibnamefont{Bernstein}}
  \bibnamefont{and} \bibinfo{editor}{\bibfnamefont{B.~R.}
  \bibnamefont{Holstein}} (\bibinfo{publisher}{Springer-Verlag},
  \bibinfo{address}{Berlin, Heidelberg, New York}, \bibinfo{year}{1995}), vol.
  \bibinfo{volume}{452} of \emph{\bibinfo{series}{Lecture Notes in Physics}}.

\end{thebibliography}
\end{document}